\documentclass[preprint,12pt,authoryear]{elsarticle}

\usepackage{lineno,hyperref}
\usepackage{amssymb}
\usepackage{braket}
\usepackage[super,compress]{cite}
\usepackage{epsfig}
\usepackage{epstopdf}
\usepackage{multirow}
\usepackage{booktabs}
\usepackage{adjustbox}
\usepackage{caption}
\usepackage{array}
\usepackage{verbatim}
\usepackage{stmaryrd}

\usepackage{boondox-cal}

\usepackage{rotating}
\usepackage{etoolbox}

\usepackage{boondox-cal}

\journal{New Astronomy}

\usepackage{numcompress}

\begin{document}

\begin{frontmatter}

\title{Lepton emission rates of $^{43-64}$V isotopes under stellar conditions}

\author[mymainaddress]{Ramoona Shehzadi\corref{mycorrespondingauthor}}
\cortext[mycorrespondingauthor]{Corresponding author}
\ead{ramoona.physics@pu.edu.pk}

\author[mysecondaryaddress]{Jameel-Un Nabi}
\ead{jameel@giki.edu.pk}

\author[mymainaddress]{Fakeha Farooq}
\ead{fakehafarooq@gmail.com}

\address[mymainaddress]{Department of Physics, University of the Punjab, Lahore 54590, Pakistan}
\address[mysecondaryaddress]{Faculty of Engineering Sciences, GIK
Institute of Engineering Sciences and Technology, Topi Khyber Pakhtunkhwa 23640, Pakistan}

\begin{abstract}
In astrophysical conditions prevalent during the late times of
stellar evolution, lepton ($e^{-}$ and $e^{+}$) emission processes
compete with the corresponding lepton capture processes. Prior to
the collapse, lepton emissions significantly affect the cooling of
the core and reduce its entropy. Therefore the lepton emission rates
for Fe-group nuclei serve as an important input for core-collapse
simulations of high-mass stars. From earlier simulation studies,
isotopes of vanadium (V) have great astrophysical significance in
regard to their weak-decay rates which substantially affect Y$_{e}$
(fraction of lepton to baryon number) during the final developmental
stages of massive stars. The current study involves the computation
of the weak lepton emission (LE) rates for V-isotopes by employing
the improved deformed proton-neutron Quasi-particle Random Phase
Approximation (pn-QRPA) model. The mass numbers of the selected
isotopes range from 43 to 64. The LE rates on these isotopes have
been estimated for a broad spectrum of density and temperature under
astrophysical conditions. The ranges considered for density and
temperature are 10$^{1}$ to 10$^{11}$ (g/cm$^{3}$) and 10$^{7}$ to
3$\times$10$^{11}$ (K), respectively. The lepton emission rates from
the present study were also compared to the rates previously
estimated by using the independent-particle model (IPM) and large-scale shell model (LSSM).
IPM rates are generally bigger than QRPA rates, while LSSM rates are overall in good comparison with the reported
rates. We attribute these differences to correct placement of GT centroid in LSSM and QRPA models.
\end{abstract}

\begin{keyword}
Lepton emission rates; Gamow-Teller transitions; core-collapse; pn-QRPA model
\end{keyword}

\end{frontmatter}


%
\section{Introduction}
At late times of hydrostatic nuclear stellar-evolution, the weak
$\beta^{\pm}$ processes, accompanying electron and positron (lepton) emissions, involving
the iron-peak nuclei play a consequential role~\citep{Burbidge57,Bethe90,Lang03,Rau02}.
These processes help to determine the pre-collapse stellar structure and the nucleosynthesis by
evolving Y$_{e}$ (fraction of number of leptons to number of baryons)
during the last stages of life track of high-mass stars~\citep{Aufder94,Jos11}.
From H-burning to the starting point of burning phase of carbon, both
lepton capture (LC) and lepton emission (LE) processes change Y$_{e}$
from 1 to $\sim0.5$ and then to around $\sim0.42$ during pre-collapse
stages~\citep{Nab11}. Just after the completion of Si-burning phase,
electron capturing dominates. When evolution proceeds, Y$_{e}$ drops
and leads to pre-supernova environment, abundant with neutron-rich nuclei
having large lepton emission rates. In astrophysical conditions prevailing prior to the
collapse, for Y$_{e}$ in the range $\sim0.42$ to $0.46$, LE rates are
strong enough to compete with the LC and at Y$_{e}\sim0.456$,
they can balance the capture rates~\citep{Auf94}. After this balancing
stage, the neutrinos produced in the LE processes have very large emission energies.
Therefore, the cooling of the core before collapse might be significant because of
LE processes. Thus, during the late evolution stages, LE rates largely
contribute in changing Y$_{e}$ and entropy of the stellar core~\citep{Janka07}.

Before 1980, many authors had performed calculations of weak rates
under stellar conditions. Various studies showed that the inclusion of
GT resonances in the evaluation of astrophysical weak-decay rates
significantly enhances the rates when compared with their previously calculated
values. In the first place, the significance of Gamow Teller (GT)
resonance for the electron captures (EC) on fp-shell nuclei in stellar
environment was suggested by Bethe and
collaborators~\citep{Bethe79}. A major breakthrough in the
calculations of nuclear astrophysical rates was made by the authors
Fuller, Fowler and Newman (FFN)~\citeyearpar{Fuller80, Fuller82a, Fuller82b, Fuller85}, when they
systematically included these resonances in their computations of
the stellar rates both in EC and $\beta^{\pm}$ directions. Their
rates were much stronger than the earlier rates calculated
by~\citet{Maz74} and~\citet{Han68} without including the resonances. They also
reported the primary role of the GT-centroid in the determination of
the effective energy of the LE and LC reactions. They used a
parametrization method as used in the independent particle model
(IPM) and computed the strength and location of GT resonances for
226 nuclei (21 $\leq$ A $\leq$ 60). Later,~\citet{Aufder94} improved the work of FFN with the addition of
GT quenching and calculated the rates for heavier nuclei having
A$>$60. The results of (n,p) and (p,n) experiments~\citep{Kateb94, Ron93, Rap83, And90, Good80} later revealed the shortcomings in the
computations done by Aufderheide et al. and FFN. The measured total
GT strength was quenched in comparison to the IPM calculated
strength and these strengths were greatly segmented over several
final decay-states of daughter nuclei. It was emphasized that the
origin of these effects was the valance nucleon-nucleon residual
interactions. Another flaw in the parametrization used by FFN and
Aufderheide et al. was the incorrect placement of GT centroid.

Afterwards, instead of using phenomenologically based
parametrization of GT strengths and centroids, theoretical efforts
were made to accurately describe the correlations among the valance
nucleons for a reliable evaluation of astrophysical weak rates on
microscopic level. Today, it is well known that the proton-neutron
quasi-particle random phase approximation (pn-QRPA)~\citep{Nabi99}
and the large-scale shell model (LSSM)~\citep{Lang00} are reliable
theories which microscopically deal with the computations of stellar
weak rates. However, the LSSM calculations used an approximation
method based on Brink's hypothesis~\citep{Brink95, Axel62} to
incorporate the contribution of GT strength distributions from
parent excitation states. According to this hypothesis, the GT
strength distributions of excited levels are displaced from that of
ground state only by an amount equal to the excitation energy of
that state. Since, the temperature conditions which exist in the
stellar matter during pre-collapse and supernova stages are so
intense ($\sim10^{9}\;$K) that the excited states of parent nuclei
have considerable occupation probability. Thus, the individual
excited states give measurable contributions to the total
stellar-weak rates. Therefore, the method based on microscopical
calculation of rates must include the contributions of all the
partial decay rates due to individual parent excited states. This
state-by-state evaluation of weak-interaction mediated rates is the
foundation of the pnQRPA model. In addition, this model makes use of
large shell model space up till 7$\mathsf{\hbar\omega}$, and
therefore can compute the rates for an arbitrarily heavy nuclide.
These features of pn-QRPA model increase its reliability and utility
in stellar-weak rates calculations. This model was employed by Nabi
and collaborators, e.g.,
in~\citep{Nabi99,Nabi04,Nabi05,Nabi07,Nabi08,Nabi17}, where they
successfully computed the weak rates of several Fe-peak nuclei
having crucial importance.

Isotopes of vanadium are considered amongst the notable iron-peak
$\beta$-decay nuclei in the pre-collapse developments of massive
stars~\citep{Aufder94, Heger01}. For temporal variation of Y$_{e}$
in the range, 0.40 $\leq$ Y$_{e}$ $\leq$ 0.5, within the core of
massive stars, the simulation studies of~\citet{Aufder94} revealed
the top 71 EC and $\beta$-decay nuclei. From these studies, the
isotopes of vanadium $^{50, 52-57}V$ and $^{50-55}V$ were
short-listed as crucial $\beta$-decay and EC nuclei respectively,
having considerable impact in stellar trajectory during and post
Si-burning stages. The weak-decay characteristics of stable isotopes
of vanadium $^{50}$V and $^{51}$V and their astrophysical
implications were focused in several studies. For example,
microscopic calculations of EC rates of $^{50}$V and $^{51}$V by
using the pnQRPA theory were performed by~\citet{Nabi07}
and~\citet{Muneeb13}, respectively. The astrophysical significance
of these isotopes was also highlighted in the work of~\citet{C05},~\citet{Cole12} and~\citet{Sarrig16}. In recent studies, Shehzadi et al. have
presented the detailed analysis of energy rates including gamma
heating and neutrino cooling~\citep{shehzadi20}, and lepton capture
weak rates~\citep{shehzadi20a} for vanadium isotopes series in the
mass range 43 $\leq$ A $\leq$ 64, by using the deformed pnQRPA
model. In the current paper, we have computed the lepton emission
(LE) rates of the vanadium isotopes by using the B(GT) data already
published in~\citet{shehzadi20}. In addition we have compared
our results to that of IPM and LSSM model rates. The formalism of pn-QRPA model
is described in the next section. Section~\ref{sec:results} involves calculated results and discussion. The last Section concludes the results.
\section{Formalism}
\label{sec:formalism}

In the calculations of the pn-QRPA theory, the system of
quasiparticles is handled by Nilsson model~\citep{Nil55} and BCS
approximation. The Nilsson model deals with the single-particle
($\mathsf{sp}$) hamiltonian ($\mathsf{H_{sp}}$) and estimates the
$\mathsf{sp}$-states and their energies. This model also takes into
account the deformation of the nucleus. Pairing nucleon-nucleon
correlations ($\mathsf{V_{pair}}$) were treated under the BCS
approximation. The residual interactions among the pairs of
proton-neutron were employed through two separate GT forces, known
as the particle-hole ($\mathsf{ph}$) and particle-particle
($\mathsf{pp}$) GT interactions, represented by
$\mathsf{V_{ph(GT)}}$ and $\mathsf{V_{pp(GT)}}$, respectively. In
the pn-QRPA theory, these interactions are specified by the force
parameters $\kappa$ (for $\mathsf{pp}$) and $\chi$ (for
$\mathsf{ph}$). For the isotopes under study, the values of these
parameters were selected in such a way that their experimentally
measured half-lives, as stated in~\citet{Aud17}, could be reproduced
by our model. The following forms of $\kappa$ and $\chi$, as
in~\citep{Hir93}, were adopted;
\begin{equation}
 \kappa = A^{-2/3} \;(MeV);~~~~~~\chi = 23 A^{-1} \;(MeV)
\end{equation}
After combining the above mentioned interaction terms in one
equation, the accumulated hamiltonian of the pn-QRPA model has the
form,
\begin{equation}
\mathsf{H_{pnQRPA}} = \mathsf{H_{sp}} + \mathsf{V_{pair}} + \mathsf{V_{pp(GT)}} + \mathsf{V_{ph(GT)}}
\end{equation}

Amongst other important parameters of the model are the pairing
gaps, whose values were set according to;
\begin{equation}
\Delta _{\mathsf{n}} = \Delta _{\mathsf{p}} = 12\mathsf{A^{-1/2}}\;(\mathsf{MeV})
\label{Eq:pairgap}
\end{equation}
The Nilsson potential (NP) parameters, which were chosen
from~\citet{Nil55} and the Nilsson oscillator constant was
calculated using the formula $\Omega =
41\mathsf{A^{-1/3}}\;(\mathsf{MeV})$. The nuclear quadrupole
deformation, $\mathcal{Q}$ was evaluated from;
\begin{equation}
\mathcal{Q} = \frac{125(\mathcal{q})}{1.44(\mathsf{A^{2/3}})(\mathsf{Z})};
\label{Eq:deform}
\end{equation}
where $\mathcal{q}$ is the electric quadruple moment taken from
~\citep{Mol95} and $\mathsf{Z}$ ($\mathsf{A}$) is the atomic number
(mass number). The latest data of mass compilation
from~\citet{Aud17} was used to calculate the reaction Q-values.

The computation of the weak-decay rates in the stellar interior was
performed by adopting the same approach as used in the previous
calculations done by~\citet{Fuller}. However, in our model GT
strength for decay rates from all excitation states were computed
microscopically. The rates of the electron and positron emission,
from parent $\mathcal{i}^{th}$ state to the $\mathcal{j}^{th}$ state
of product nuclide, were computed as;
\begin{eqnarray}
\lambda ^{LE} _{\mathcal{ij}} &=& \left(\frac{\ln 2}{\mathcal{D}} \right)
[f_{\mathcal{ij}} (T, E_{f}, \rho)][B(F)_{\mathcal{ij}} \nonumber \\
&+&(\mathcal{g}_{A}/ \mathcal{g}_{V})^{2} B(GT)_{\mathcal{ij}}];~~~~ L\equiv E,P\label{Eq:LE}.
\end{eqnarray}

The values of $\mathcal{D}$  and $\mathcal{g}_{A}/ \mathcal{g}_{V}$ are 6143~\citep{Har09}
and -1.2694~\citep{Nak10}, respectively. B(F)$_{\mathcal{ij}}$ and B(GT)$_{\mathcal{ij}}$ in Eq.~\ref{Eq:LE}, are
reduced Fermi and GT transition probabilities, respectively which are given as;
\begin{eqnarray}
B(F)_{\mathcal{ij}} &=& \frac{\mathcal{1}}{2J_{i}+1}
|\langle \mathcal{j}||\mathcal{\sum_{l}}t^{\mathcal{l}}_{\pm}||\mathcal{i \rangle|^{2}}\\
B(GT)_{\mathcal{ij}} &=& \frac{\mathtt{1}}{2J_{i}+1}
|\langle \mathcal{j}||\mathcal{\sum_{l}}t^{\mathcal{l}}_{\pm}\vec{\sigma}^{\mathcal{l}}||\mathcal{i \rangle|^{2}}
\label{Eq:FGTP}
\end{eqnarray}
where $J_{i}$, $\vec{\sigma}^{\mathcal{l}}$ and $t^{\mathcal{l}}_{\pm}$ are the
total $\mathcal{i}^{th}$ state spin of nucleus, Pauli spin operator and raising (lowering) isospin operators, respectively.
The isospin and spin operators act on the $\mathcal{l}^{th}$ nucleon of a nucleus.

In Eq.~\ref{Eq:LE}, $f_{\mathcal{ij}}$, is the phase space integral (in natural units) which is carried over total energy.
This integral has the following form for the lepton emission rates (with lower sign for PE and upper one for EE);
\begin{eqnarray}
f_{\mathcal{ij}} &=& \int _{1}^{\mathcal{w}_{\mathcal{j}}}\mathcal{w}
(\mathcal{w}^{2} -1)^{1/2}(\mathcal{w}_{\mathcal{j}}-\mathcal{w})^{3} \nonumber \\
&~&F(\pm Z, \mathcal{w})(1-G_{\mp })d\mathcal{w}\label{phemission}.
\end{eqnarray}
In this equation, $\mathcal{w}$ is the total energy, which includes the kinetic energy
and rest mass energy of the lepton. The total energy of $\beta$-decay is represented by $\mathcal{w}_{\mathcal{j}}$.
The Fermi functions  $F\left(\pm Z,\mathcal{w}\right)$ are
derived using the method similar to~\citep{Gove71}. The G$_{\mp}$ are the Fermi-Dirac distribution functions of leptons.

The total lepton emission rates are determined by;
\begin{equation}
\lambda^{LE} = \sum _{\mathcal{ij}}P_{\mathcal{i}} \lambda _{\mathcal{ij}}^{LE};~~~~ L\equiv E,P,
\label{rate}
\end{equation}
where $P_{\mathcal{i}}$ is the probability of occupation of the
parent excitation levels obeying the normal Boltzmann distribution.
The sum in Equation~\ref{rate} is applied over a set of all those
levels in parent and daughter nuclei, which result in the
convergence of the calculated rates.

\section{Results and Discussions} \label{sec:results}
In this study, the calculations of lepton emission rates on vanadium
isotopes have peen performed by utilizing the deformed pn-QRPA
model. The selected vanadium isotopes have mass numbers from A =
43-64. Out of these isotopes, $^{50,51}$V are stable and the
remaining ones are unstable including neutron-deficit and
neutron-rich nuclide. A broad domain of stellar temperature, $10^{7}
- 3 \times 10^{10}\;$K, and density, $10-10^{11}\;$g/cm$^{3}$, has
been considered for the fine grid calculations of the rates. The
lepton emission rates from current calculations have also been
compared to the corresponding rates computed by the IPM and LSSM
models.

Tables~$\ref{V47-52}-\ref{V59-64}$ and~\ref{V43-48}-\ref{V49-54} show the EE
and PE rates of $^{47-64}$V and $^{43-54}$V isotopes, respectively. In case of
neutron deficit nuclei ($^{43-46}$V), the calculated EE rates and for neutron
rich nuclei ($^{55-64}$V), PE rates are smaller than $10^{-100}\;$s$^{-1}$
and hence have not been shown here. First two columns of each table present
the chosen values of density, $\rho Y_{e}$ (in units of g/cm$^{3}$) and
temperature, T$_{9}$ (in 10$^{9}\;$K), respectively. Because of the space
limitations, the rates calculated only at some specific values within the selected
ranges of temperature and density are shown here. The remaining columns of
the tables show the calculated EE ($\lambda^{EE}$) or PE ($\lambda^{PE}$)
weak-decay rates both in s$^{-1}$ unit. From
Tables~\ref{V47-52}-\ref{V59-64}, it can be seen that, at some given density
and temperature, the EE rates for $^{64}$V are strongest while for $^{47}$V
are weakest. The tables show that for every isotope, overall EE rates in each
density region i.e., low (10$^{2}\;$g/cm$^{3}$), medium (10$^{5}$,
10$^{8}\;$g/cm$^{3}$) and high (10$^{11}\;$g/cm$^{3}$), increase with
increasing temperature by several orders of magnitudes. This happens since
the weak rates largely depend on the available phase space which shows
considerable expansion with temperature. In addition, with increasing
temperature, the occupation probabilities of parent excited states enhance
which largely contribute in increasing the values of total rates. One can
also notice that, for a specific value of temperature, as the density of the
core increases from 10$^{1}$ to 10$^{6}\;$g/cm$^{3}$, the EE rates do not
change considerably. However, with a further increase in the core density,
the EE rates start to decrease. This occurs, because with increment in core
density it becomes stiff and the available phase space reduces which weakens
the EE rates appreciably.

\begin{table}[pt]
\caption{The electron emission rates, $\lambda^{EE}$ (in s$^{-1}$), calculated
using pn-QRPA model for $^{47-52}$V isotopes at different values of stellar
densities, $\rho$Y$_{e}$, and temperatures, T$_{9}$,. $\rho$Y$_{e}$ has units of g/cm$^{3}$, where
Y$_{e}$ is the ratio of number of leptons to number of baryons and
$\rho$ is the baryon density. T$_{9}$ is given in units of $10^{9}\;$K. }\label{V47-52} \centering
\begin{tabular}{cccccccc}
\toprule
{$\rho$Y$_{e}$} &
{T$_{9}$} & {$^{47}$V}&
{$^{48}$V} & {$^{49}$V}&
{$^{50}$V}& {$^{51}$V}& {$^{52}$V}\\
\midrule
10$^{2}$ & 1 & 2.93E-46 & 1.74E-33 & 2.86E-27 & 8.20E-21 & 1.15E-23 & 2.75E-03\tabularnewline
10$^2$ & 1.5 & 1.16E-33 & 1.41E-23 & 7.35E-21 & 1.31E-15 & 8.05E-17 & 3.10E-03\tabularnewline
10$^2$ & 2 & 3.48E-27 & 1.31E-18 & 1.95E-16 & 1.63E-12 & 2.29E-13 & 5.43E-03\tabularnewline
10$^2$ & 3 & 4.06E-20 & 1.21E-13 & 6.38E-12 & 2.46E-09 & 7.11E-10 & 1.75E-02\tabularnewline
10$^2$ & 5 & 3.10E-14 & 1.07E-09 & 2.50E-08 & 1.44E-06 & 7.13E-07 & 5.15E-02\tabularnewline
10$^2$ & 10 & 7.01E-10 & 9.20E-07 & 1.25E-05 & 3.27E-04 & 7.23E-04 & 1.30E-01\tabularnewline
10$^2$ & 30 & 2.78E-07 & 9.20E-05 & 7.45E-04 & 1.56E-02 & 1.23E-01 & 1.82E+00\tabularnewline
 &  &  &  &  &  &  & \tabularnewline
10$^5$ & 1 & 2.33E-46 & 1.60E-33 & 2.38E-27 & 7.03E-21 & 1.14E-23 & 2.54E-03\tabularnewline
10$^5$  & 1.5 & 1.05E-33 & 1.35E-23 & 7.19E-21 & 1.29E-15 & 8.04E-17 & 2.98E-03\tabularnewline
10$^5$  & 2 & 3.37E-27 & 1.28E-18 & 1.94E-16 & 1.61E-12 & 2.29E-13 & 5.36E-03\tabularnewline
10$^5$  & 3 & 4.05E-20 & 1.20E-13 & 6.37E-12 & 2.45E-09 & 7.10E-10 & 1.74E-02\tabularnewline
10$^5$  & 5 & 3.09E-14 & 1.07E-09 & 2.50E-08 & 1.44E-06 & 7.13E-07 & 5.14E-02\tabularnewline
10$^5$  & 10 & 7.01E-10 & 9.20E-07 & 1.25E-05 & 3.27E-04 & 7.23E-04 & 1.30E-01\tabularnewline
10$^5$  & 30 & 2.78E-07 & 9.20E-05 & 7.45E-04 & 1.56E-02 & 1.23E-01 & 1.82E+00\tabularnewline
 &  &  &  &  &  &  & \tabularnewline
10$^8$ & 1 & 1.52E-54 & 6.61E-41 & 4.42E-31 & 6.87E-27 & 3.01E-24 & 1.14E-10\tabularnewline
10$^8$ & 1.5 & 2.38E-38 & 1.32E-28 & 6.19E-22 & 4.33E-19 & 2.30E-17 & 1.02E-07\tabularnewline
10$^8$ & 2 & 3.57E-30 & 2.17E-22 & 2.47E-17 & 4.61E-15 & 7.01E-14 & 3.80E-06\tabularnewline
10$^8$ & 3 & 6.58E-22 & 4.23E-16 & 1.11E-12 & 7.05E-11 & 2.51E-10 & 1.81E-04\tabularnewline
10$^8$ & 5 & 3.31E-15 & 5.43E-11 & 7.48E-09 & 2.56E-07 & 3.54E-07 & 5.07E-03\tabularnewline
10$^8$ & 10 & 3.64E-10 & 4.27E-07 & 8.32E-06 & 1.99E-04 & 5.24E-04 & 7.35E-02\tabularnewline
10$^8$ & 30 & 2.69E-07 & 8.91E-05 & 7.28E-04 & 1.51E-02 & 1.20E-01 & 1.77E+00\tabularnewline
 &  &  &  &  &  &  & \tabularnewline
10$^{11}$ & 1 & 1.00E-100 & 1.00E-100 & 1.00E-100 & 1.00E-100 & 1.00E-100 & 1.00E-100\tabularnewline
10$^{11}$ & 1.5 & 1.00E-100 & 1.00E-100 & 9.93E-93 & 6.82E-91 & 5.62E-86 & 9.82E-80\tabularnewline
10$^{11}$ & 2 & 2.19E-84 & 1.25E-76 & 1.77E-70 & 7.03E-69 & 2.99E-65 & 3.85E-60\tabularnewline
10$^{11}$ & 3 & 4.21E-58 & 2.48E-52 & 4.13E-48 & 9.93E-47 & 2.20E-44 & 1.81E-40\tabularnewline
10$^{11}$ & 5 & 5.25E-37 & 7.67E-33 & 4.61E-30 & 8.00E-29 & 1.69E-27 & 1.19E-24\tabularnewline
10$^{11}$ & 10 & 3.67E-21 & 4.19E-18 & 2.33E-16 & 3.42E-15 & 2.00E-14 & 1.14E-12\tabularnewline
10$^{11}$ & 30 & 9.86E-11 & 3.50E-08 & 3.75E-07 & 6.58E-06 & 5.71E-05 & 8.22E-04\tabularnewline
\bottomrule
\end{tabular}
\end{table}

\begin{table}[pt]
\caption{Same as in Table~\ref{V47-52}, but for electron emission
rates, $\lambda^{EE}$, due to $^{53-58}$V.}\label{V53-58} \centering
\begin{tabular}{cccccccc}
\toprule
{$\rho$Y$_{e}$} &
{T$_{9}$} & {$^{53}$V}&
{$^{54}$V} & {$^{55}$V}&
{$^{56}$V}& {$^{57}$V}& {$^{58}$V}\\
\midrule
10$^{2}$ & 1 & 6.95E-03 & 1.48E-02 & 7.18E-02 & 2.60E+00 & 1.47E+00 & 5.73E+00\tabularnewline
10$^{2}$ & 1.5 & 6.85E-03 & 1.77E-02 & 6.89E-02 & 2.48E+00 & 1.42E+00 & 9.31E+00\tabularnewline
10$^{2}$ & 2 & 6.76E-03 & 2.20E-02 & 6.59E-02 & 2.39E+00 & 1.36E+00 & 1.26E+01\tabularnewline
10$^{2}$ & 3 & 6.52E-03 & 3.67E-02 & 6.03E-02 & 2.29E+00 & 1.26E+00 & 1.72E+01\tabularnewline
10$^{2}$ & 5 & 6.18E-03 & 1.02E-01 & 5.19E-02 & 2.49E+00 & 1.17E+00 & 2.25E+01\tabularnewline
10$^{2}$ & 10 & 2.57E-02 & 4.51E-01 & 8.81E-02 & 4.55E+00 & 7.13E+00 & 2.84E+01\tabularnewline
10$^{2}$ & 30 & 1.87E+00 & 4.19E+00 & 2.42E+00 & 2.77E+01 & 1.07E+02 & 9.75E+01\tabularnewline
 &  &  &  &  &  &  & \tabularnewline
10$^{5}$ & 1 & 6.92E-03 & 1.47E-02 & 7.16E-02 & 2.59E+00 & 1.47E+00 & 5.73E+00\tabularnewline
10$^{5}$ & 1.5 & 6.84E-03 & 1.77E-02 & 6.89E-02 & 2.48E+00 & 1.42E+00 & 9.29E+00\tabularnewline
10$^{5}$ & 2 & 6.75E-03 & 2.19E-02 & 6.59E-02 & 2.39E+00 & 1.36E+00 & 1.26E+01\tabularnewline
10$^{5}$ & 3 & 6.50E-03 & 3.66E-02 & 6.03E-02 & 2.29E+00 & 1.26E+00 & 1.72E+01\tabularnewline
10$^{5}$ & 5 & 6.17E-03 & 1.02E-01 & 5.19E-02 & 2.49E+00 & 1.17E+00 & 2.25E+01\tabularnewline
10$^{5}$ & 10 & 2.57E-02 & 4.51E-01 & 8.81E-02 & 4.55E+00 & 7.13E+00 & 2.84E+01\tabularnewline
10$^{5}$ & 30 & 1.87E+00 & 4.19E+00 & 2.42E+00 & 2.77E+01 & 1.07E+02 & 9.75E+01\tabularnewline
 &  &  &  &  &  &  & \tabularnewline
10$^{8}$ & 1 & 1.46E-03 & 4.56E-03 & 4.97E-02 & 1.72E+00 & 1.19E+00 & 4.78E+00\tabularnewline
10$^{8}$ & 1.5 & 1.51E-03 & 5.79E-03 & 4.80E-02 & 1.65E+00 & 1.15E+00 & 7.74E+00\tabularnewline
10$^{8}$ & 2 & 1.57E-03 & 7.82E-03 & 4.60E-02 & 1.60E+00 & 1.11E+00 & 1.05E+01\tabularnewline
10$^{8}$ & 3 & 1.73E-03 & 1.65E-02 & 4.27E-02 & 1.56E+00 & 1.03E+00 & 1.44E+01\tabularnewline
10$^{8}$ & 5 & 2.29E-03 & 6.27E-02 & 3.84E-02 & 1.80E+00 & 9.75E-01 & 1.91E+01\tabularnewline
10$^{8}$ & 10 & 2.03E-02 & 3.72E-01 & 7.59E-02 & 3.85E+00 & 6.50E+00 & 2.58E+01\tabularnewline
10$^{8}$ & 30 & 1.84E+00 & 4.11E+00 & 2.37E+00 & 2.72E+01 & 1.06E+02 & 9.64E+01\tabularnewline
 &  &  &  &  &  &  & \tabularnewline
10$^{11}$ & 1 & 1.00E-100 & 1.00E-100 & 1.87E-93 & 1.38E-92 & 1.01E-81 & 1.97E-79\tabularnewline
10$^{11}$ & 1.5 & 4.12E-72 & 1.72E-69 & 1.24E-63 & 1.43E-62 & 1.69E-55 & 1.10E-53\tabularnewline
10$^{11}$ & 2 & 5.43E-55 & 1.19E-52 & 1.25E-48 & 1.87E-47 & 2.76E-42 & 1.05E-40\tabularnewline
10$^{11}$ & 3 & 9.59E-38 & 1.15E-35 & 1.69E-33 & 3.58E-32 & 6.15E-29 & 1.46E-27\tabularnewline
10$^{11}$ & 5 & 9.64E-24 & 7.50E-22 & 3.49E-21 & 1.16E-19 & 5.01E-18 & 8.81E-17\tabularnewline
10$^{11}$ & 10 & 1.91E-12 & 3.61E-11 & 2.42E-11 & 7.57E-10 & 6.15E-09 & 2.88E-08\tabularnewline
10$^{11}$ & 30 & 1.21E-03 & 2.79E-03 & 2.07E-03 & 2.23E-02 & 1.14E-01 & 1.27E-01\tabularnewline
\bottomrule
\end{tabular}
\end{table}

\begin{table}[pt]
\caption{Same as in Table~\ref{V47-52} but for electron emission rates, $\lambda^{EE}$,
due to $^{59-64}$V.}\label{V59-64}
\centering
\begin{tabular}{cccccccc}
\toprule
{$\rho$Y$_{e}$} &
{T$_{9}$} & {$^{59}$V}&
{$^{60}$V} & {$^{61}$V}&
{$^{62}$V}& {$^{63}$V}& {$^{64}$V}\\
\midrule
10$^{2}$ & 1 & 6.35E+00 & 5.82E+00 & 1.42E+01 & 5.04E+01 & 5.28E+01 & 5.98E+01\tabularnewline
10$^{2}$ & 1.5 & 6.17E+00 & 6.64E+00 & 1.47E+01 & 6.41E+01 & 5.85E+01 & 8.57E+01\tabularnewline
10$^{2}$ & 2 & 6.07E+00 & 7.69E+00 & 1.53E+01 & 7.13E+01 & 6.17E+01 & 1.14E+02\tabularnewline
10$^{2}$ & 3 & 5.90E+00 & 9.68E+00 & 1.67E+01 & 7.85E+01 & 6.50E+01 & 1.60E+02\tabularnewline
10$^{2}$ & 5 & 6.03E+00 & 1.24E+01 & 1.90E+01 & 8.49E+01 & 6.82E+01 & 2.04E+02\tabularnewline
10$^{2}$ & 10 & 8.61E+01 & 1.66E+01 & 2.64E+01 & 1.09E+02 & 9.27E+01 & 2.61E+02\tabularnewline
10$^{2}$ & 30 & 2.07E+03 & 7.82E+01 & 2.08E+02 & 5.04E+02 & 5.52E+02 & 1.26E+03\tabularnewline
 &  &  &  &  &  &  & \tabularnewline
10$^{5}$ & 1 & 6.34E+00 & 5.82E+00 & 1.42E+01 & 5.04E+01 & 5.28E+01 & 5.98E+01\tabularnewline
10$^{5}$ & 1.5 & 6.17E+00 & 6.64E+00 & 1.47E+01 & 6.41E+01 & 5.85E+01 & 8.57E+01\tabularnewline
10$^{5}$ & 2 & 6.07E+00 & 7.69E+00 & 1.53E+01 & 7.13E+01 & 6.17E+01 & 1.14E+02\tabularnewline
10$^{5}$ & 3 & 5.90E+00 & 9.68E+00 & 1.67E+01 & 7.85E+01 & 6.50E+01 & 1.60E+02\tabularnewline
10$^{5}$ & 5 & 6.03E+00 & 1.24E+01 & 1.90E+01 & 8.49E+01 & 6.82E+01 & 2.04E+02\tabularnewline
10$^{5}$ & 10 & 8.61E+01 & 1.66E+01 & 2.64E+01 & 1.09E+02 & 9.27E+01 & 2.61E+02\tabularnewline
10$^{5}$ & 30 & 2.07E+03 & 7.82E+01 & 2.08E+02 & 5.04E+02 & 5.52E+02 & 1.26E+03\tabularnewline
 &  &  &  &  &  &  & \tabularnewline
10$^{8}$ & 1 & 5.36E+00 & 5.18E+00 & 1.26E+01 & 4.61E+01 & 4.81E+01 & 5.38E+01\tabularnewline
10$^{8}$ & 1.5 & 5.21E+00 & 5.92E+00 & 1.30E+01 & 5.87E+01 & 5.33E+01 & 7.71E+01\tabularnewline
10$^{8}$ & 2 & 5.13E+00 & 6.85E+00 & 1.36E+01 & 6.55E+01 & 5.62E+01 & 1.03E+02\tabularnewline
10$^{8}$ & 3 & 5.00E+00 & 8.63E+00 & 1.49E+01 & 7.23E+01 & 5.94E+01 & 1.45E+02\tabularnewline
10$^{8}$ & 5 & 5.18E+00 & 1.11E+01 & 1.72E+01 & 7.85E+01 & 6.28E+01 & 1.87E+02\tabularnewline
10$^{8}$ & 10 & 8.11E+01 & 1.56E+01 & 2.48E+01 & 1.03E+02 & 8.79E+01 & 2.47E+02\tabularnewline
10$^{8}$ & 30 & 2.05E+03 & 7.73E+01 & 2.06E+02 & 4.99E+02 & 5.47E+02 & 1.24E+03\tabularnewline
 &  &  &  &  &  &  & \tabularnewline
10$^{11}$ & 1 & 4.50E-71 & 6.61E-71 & 1.75E-62 & 8.09E-60 & 6.04E-52 & 4.79E-52\tabularnewline
10$^{11}$ & 1.5 & 1.95E-48 & 4.02E-48 & 1.11E-42 & 1.17E-40 & 8.22E-36 & 2.07E-35\tabularnewline
10$^{11}$ & 2 & 5.57E-37 & 1.33E-36 & 1.17E-32 & 5.87E-31 & 1.23E-27 & 5.77E-27\tabularnewline
10$^{11}$ & 3 & 2.51E-25 & 6.24E-25 & 1.73E-22 & 4.32E-21 & 2.61E-19 & 2.31E-18\tabularnewline
10$^{11}$ & 5 & 1.08E-15 & 2.29E-15 & 3.92E-14 & 6.14E-13 & 2.07E-12 & 2.94E-11\tabularnewline
10$^{11}$ & 10 & 4.00E-07 & 8.36E-08 & 2.71E-07 & 3.23E-06 & 1.90E-06 & 1.73E-05\tabularnewline
10$^{11}$ & 30 & 3.32E+00 & 1.29E-01 & 4.09E-01 & 1.48E+00 & 1.17E+00 & 3.26E+00\tabularnewline
\bottomrule
\end{tabular}
\end{table}

\begin{table}[pt]
\caption{The positron emission rates, $\lambda^{PE}$, calculated using
pn-QRPA model for $^{43-48}$V isotopes at various selected temperatures and densities
in stellar matter. Other details same as in Table~\ref{V47-52}}\label{V43-48}
\centering
\begin{tabular}{cccccccc}
\toprule
{$\rho$Y$_{e}$} &
{T$_{9}$} & {$^{43}$V}&
{$^{44}$V} & {$^{45}$V}&
{$^{46}$V}& {$^{47}$V}& {$^{48}$V}\\
\midrule
10$^{2}$ & 1 & 1.26E+01 & 5.61E+00 & 1.95E+00 & 1.90E+00 & 3.37E-04 & 1.95E-17\tabularnewline
10$^{2}$ & 1.5 & 1.41E+01 & 5.56E+00 & 2.02E+00 & 2.26E+00 & 3.26E-04 & 4.09E-13\tabularnewline
10$^{2}$ & 2 & 1.50E+01 & 5.78E+00 & 2.05E+00 & 2.50E+00 & 3.25E-04 & 6.78E-11\tabularnewline
10$^{2}$ & 3 & 1.58E+01 & 6.92E+00 & 2.07E+00 & 2.72E+00 & 3.48E-04 & 1.35E-08\tabularnewline
10$^{2}$ & 5 & 1.62E+01 & 1.01E+01 & 2.04E+00 & 2.72E+00 & 4.39E-04 & 7.41E-06\tabularnewline
10$^{2}$ & 10 & 1.66E+01 & 2.01E+01 & 2.09E+00 & 2.22E+00 & 2.05E-03 & 5.61E-03\tabularnewline
10$^{2}$ & 30 & 6.81E+01 & 1.93E+02 & 8.79E+00 & 3.72E+00 & 1.57E-01 & 2.92E-01\tabularnewline
 &  &  &  &  &  &  & \tabularnewline
10$^{5}$ & 1 & 1.26E+01 & 5.61E+00 & 1.95E+00 & 1.90E+00 & 3.37E-04 & 1.95E-17\tabularnewline
10$^{5}$ & 1.5 & 1.41E+01 & 5.56E+00 & 2.02E+00 & 2.26E+00 & 3.26E-04 & 4.11E-13\tabularnewline
10$^{5}$ & 2 & 1.50E+01 & 5.78E+00 & 2.05E+00 & 2.50E+00 & 3.25E-04 & 6.81E-11\tabularnewline
10$^{5}$ & 3 & 1.58E+01 & 6.92E+00 & 2.07E+00 & 2.72E+00 & 3.48E-04 & 1.35E-08\tabularnewline
10$^{5}$ & 5 & 1.62E+01 & 1.01E+01 & 2.04E+00 & 2.72E+00 & 4.39E-04 & 7.41E-06\tabularnewline
10$^{5}$ & 10 & 1.66E+01 & 2.01E+01 & 2.09E+00 & 2.22E+00 & 2.05E-03 & 5.62E-03\tabularnewline
10$^{5}$ & 30 & 6.81E+01 & 1.93E+02 & 8.79E+00 & 3.72E+00 & 1.57E-01 & 2.92E-01\tabularnewline
 &  &  &  &  &  &  & \tabularnewline
10$^{8}$ & 1 & 1.26E+01 & 5.61E+00 & 1.95E+00 & 1.90E+00 & 3.37E-04 & 1.95E-17\tabularnewline
10$^{8}$ & 1.5 & 1.41E+01 & 5.56E+00 & 2.02E+00 & 2.26E+00 & 3.26E-04 & 4.11E-13\tabularnewline
10$^{8}$ & 2 & 1.50E+01 & 5.78E+00 & 2.05E+00 & 2.51E+00 & 3.26E-04 & 6.82E-11\tabularnewline
10$^{8}$ & 3 & 1.58E+01 & 6.92E+00 & 2.07E+00 & 2.75E+00 & 3.52E-04 & 1.38E-08\tabularnewline
10$^{8}$ & 5 & 1.62E+01 & 1.02E+01 & 2.06E+00 & 2.85E+00 & 4.60E-04 & 7.57E-06\tabularnewline
10$^{8}$ & 10 & 1.68E+01 & 2.05E+01 & 2.17E+00 & 2.55E+00 & 2.22E-03 & 5.87E-03\tabularnewline
10$^{8}$ & 30 & 6.87E+01 & 1.95E+02 & 8.91E+00 & 3.78E+00 & 1.59E-01 & 2.98E-01\tabularnewline
 &  &  &  &  &  &  & \tabularnewline
10$^{11}$ & 1 & 1.26E+01 & 5.61E+00 & 1.95E+00 & 1.90E+00 & 3.37E-04 & 1.95E-17\tabularnewline
10$^{11}$ & 1.5 & 1.41E+01 & 5.56E+00 & 2.02E+00 & 2.26E+00 & 3.26E-04 & 4.11E-13\tabularnewline
10$^{11}$ & 2 & 1.50E+01 & 5.78E+00 & 2.05E+00 & 2.51E+00 & 3.26E-04 & 6.82E-11\tabularnewline
10$^{11}$ & 3 & 1.58E+01 & 6.92E+00 & 2.07E+00 & 2.75E+00 & 3.52E-04 & 1.38E-08\tabularnewline
10$^{11}$ & 5 & 1.62E+01 & 1.02E+01 & 2.06E+00 & 2.85E+00 & 4.60E-04 & 7.57E-06\tabularnewline
10$^{11}$ & 10 & 1.69E+01 & 2.06E+01 & 2.18E+00 & 2.64E+00 & 2.26E-03 & 5.94E-03\tabularnewline
10$^{11}$ & 30 & 7.82E+01 & 2.25E+02 & 1.07E+01 & 4.89E+00 & 1.99E-01 & 3.86E-01\tabularnewline
\bottomrule
\end{tabular}
\end{table}

\begin{table}[pt]
\caption{Same as in Table~\ref{V43-48} but for positron emission rates, $\lambda^{PE}$,
due to $^{49-54}$V.}\label{V49-54}
\centering
\begin{tabular}{cccccccc}
\toprule
{$\rho$Y$_{e}$} &
{T$_{9}$} & {$^{49}$V}&
{$^{50}$V} & {$^{51}$V}&
{$^{52}$V}& {$^{53}$V}& {$^{54}$V}\\
\midrule
10$^{2}$ & 1 & 1.85E-11 & 5.81E-22 & 4.93E-25 & 7.64E-48 & 9.16E-40 & 5.16E-55\tabularnewline
10$^{2}$ & 1.5 & 3.57E-10 & 4.55E-17 & 2.96E-18 & 2.86E-33 & 7.98E-29 & 2.84E-38\tabularnewline
10$^{2}$ & 2 & 6.27E-09 & 4.86E-14 & 7.36E-15 & 5.55E-26 & 5.75E-23 & 6.22E-30\tabularnewline
10$^{2}$ & 3 & 2.03E-07 & 8.05E-11 & 1.86E-11 & 1.12E-18 & 1.39E-16 & 1.20E-21\tabularnewline
10$^{2}$ & 5 & 5.60E-06 & 9.77E-08 & 1.52E-08 & 7.87E-13 & 2.02E-11 & 4.04E-15\tabularnewline
10$^{2}$ & 10 & 7.00E-04 & 1.75E-04 & 1.36E-05 & 1.57E-08 & 1.19E-07 & 2.14E-10\tabularnewline
10$^{2}$ & 30 & 3.97E-02 & 3.04E-02 & 1.15E-03 & 6.14E-06 & 2.00E-05 & 1.46E-07\tabularnewline
 &  &  &  &  &  &  & \tabularnewline
10$^{5}$ & 1 & 1.85E-11 & 5.82E-22 & 4.94E-25 & 7.64E-48 & 9.16E-40 & 5.16E-55\tabularnewline
10$^{5}$ & 1.5 & 3.60E-10 & 4.56E-17 & 2.97E-18 & 2.86E-33 & 8.02E-29 & 2.85E-38\tabularnewline
10$^{5}$ & 2 & 6.28E-09 & 4.89E-14 & 7.43E-15 & 5.56E-26 & 5.78E-23 & 6.27E-30\tabularnewline
10$^{5}$ & 3 & 2.03E-07 & 8.07E-11 & 1.87E-11 & 1.12E-18 & 1.40E-16 & 1.21E-21\tabularnewline
10$^{5}$ & 5 & 5.61E-06 & 9.79E-08 & 1.52E-08 & 7.89E-13 & 2.02E-11 & 4.06E-15\tabularnewline
10$^{5}$ & 10 & 7.00E-04 & 1.75E-04 & 1.36E-05 & 1.57E-08 & 1.20E-07 & 2.14E-10\tabularnewline
10$^{5}$ & 30 & 3.97E-02 & 3.04E-02 & 1.15E-03 & 6.14E-06 & 2.00E-05 & 1.46E-07\tabularnewline
 &  &  &  &  &  &  & \tabularnewline
10$^{8}$ & 1 & 1.85E-11 & 5.82E-22 & 4.94E-25 & 7.64E-48 & 9.16E-40 & 5.16E-55\tabularnewline
10$^{8}$ & 1.5 & 3.60E-10 & 4.56E-17 & 2.97E-18 & 2.86E-33 & 8.02E-29 & 2.85E-38\tabularnewline
10$^{8}$ & 2 & 6.31E-09 & 4.90E-14 & 7.48E-15 & 5.58E-26 & 5.81E-23 & 6.31E-30\tabularnewline
10$^{8}$ & 3 & 2.07E-07 & 8.24E-11 & 1.96E-11 & 1.15E-18 & 1.42E-16 & 1.26E-21\tabularnewline
10$^{8}$ & 5 & 5.94E-06 & 1.02E-07 & 1.69E-08 & 8.69E-13 & 2.18E-11 & 4.69E-15\tabularnewline
10$^{8}$ & 10 & 7.59E-04 & 1.86E-04 & 1.52E-05 & 1.94E-08 & 1.44E-07 & 2.79E-10\tabularnewline
10$^{8}$ & 30 & 4.05E-02 & 3.10E-02 & 1.18E-03 & 6.34E-06 & 2.07E-05 & 1.51E-07\tabularnewline
 &  &  &  &  &  &  & \tabularnewline
10$^{11}$ & 1 & 1.85E-11 & 5.82E-22 & 4.94E-25 & 7.64E-48 & 9.16E-40 & 5.16E-55\tabularnewline
10$^{11}$ & 1.5 & 3.60E-10 & 4.56E-17 & 2.97E-18 & 2.86E-33 & 8.02E-29 & 2.85E-38\tabularnewline
10$^{11}$ & 2 & 6.31E-09 & 4.90E-14 & 7.48E-15 & 5.58E-26 & 5.81E-23 & 6.31E-30\tabularnewline
10$^{11}$ & 3 & 2.07E-07 & 8.24E-11 & 1.96E-11 & 1.15E-18 & 1.42E-16 & 1.26E-21\tabularnewline
10$^{11}$ & 5 & 5.94E-06 & 1.02E-07 & 1.69E-08 & 8.71E-13 & 2.18E-11 & 4.70E-15\tabularnewline
10$^{11}$ & 10 & 7.73E-04 & 1.89E-04 & 1.56E-05 & 2.04E-08 & 1.50E-07 & 2.99E-10\tabularnewline
10$^{11}$ & 30 & 5.43E-02 & 4.14E-02 & 1.71E-03 & 1.02E-05 & 3.26E-05 & 2.54E-07\tabularnewline
\bottomrule
\end{tabular}
\end{table}

The results of Tables~\ref{V43-48} and \ref{V49-54} depict that the PE rates
also increase with temperature because of an increase in the phase space.
With the rise of temperature, when degeneracy parameter of positron
becomes negative, an increasing number of positrons with higher energies are
generated resulting in larger PE rates. However, if viewed as a function of
the core density, there is no considerable change in PE rates in all density
regions. The PE rates are largest for $^{43}$V and smallest for $^{54}$V. The
comparison of PE rate tables with the corresponding EE rate tables shows
that, the PE rates for $^{47-49}$V are bigger by several orders of magnitudes
in contrast to corresponding EE rates. For next two isotopes ($^{50}$V and
$^{51}$V), from low to medium density domain, the EE rates become comparable
to the corresponding PE rates. However, in the region of high density, the PE
rates are still bigger. In case of $^{52-54}$V, the PE rates are shorter by
several orders of magnitude as compared to the corresponding EE rates. This
is because, the probability of PE processes becomes low in neutron-rich
nuclei. However, at high density $10^{11}\;$g/cm$^{3}$) PE rates again
prevail the corresponding EE rates, except at T$_{9}$ = 30. The LE rates for
V-isotopes calculated at fine density-temperature scale are available and may
be requested from corresponding author.
\begin{figure}
\begin{center}
\includegraphics[height = 1\textwidth, width=1\textwidth]{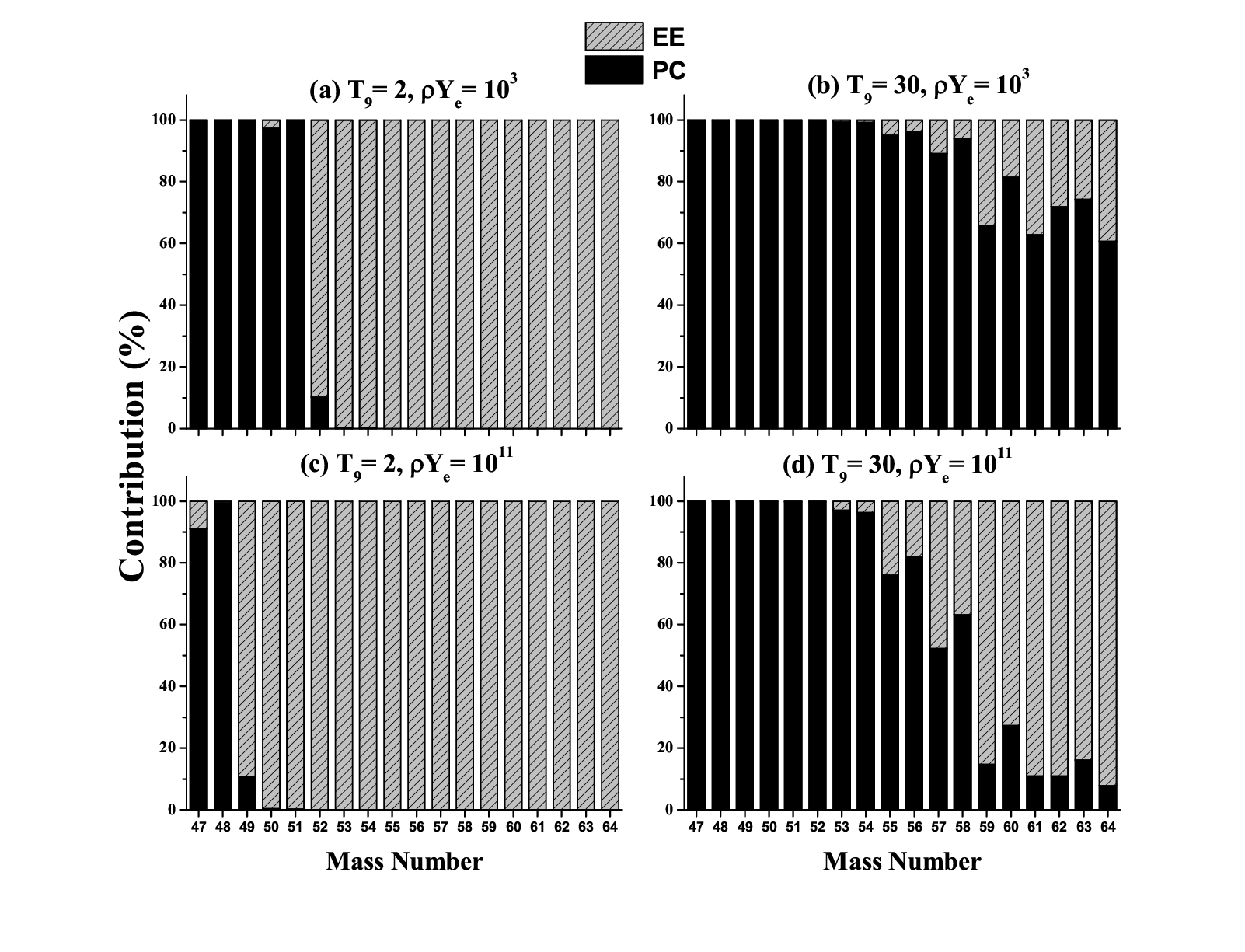}
\vspace{-0.1cm}\caption{Percentage contribution of EE and PC rates to the sum total of weak rates in $\beta^{-}$ direction at different stellar densities, $\rho$Y$_{e}$ (in g/cm$^{3}$), and temperatures, T$_{9}$ (in $10^{9}\;$K).}
\label{EE_PC_Cont}
\end{center}
\end{figure}

\begin{figure}
\begin{center}
\includegraphics[height = 1\textwidth, width=1\textwidth]{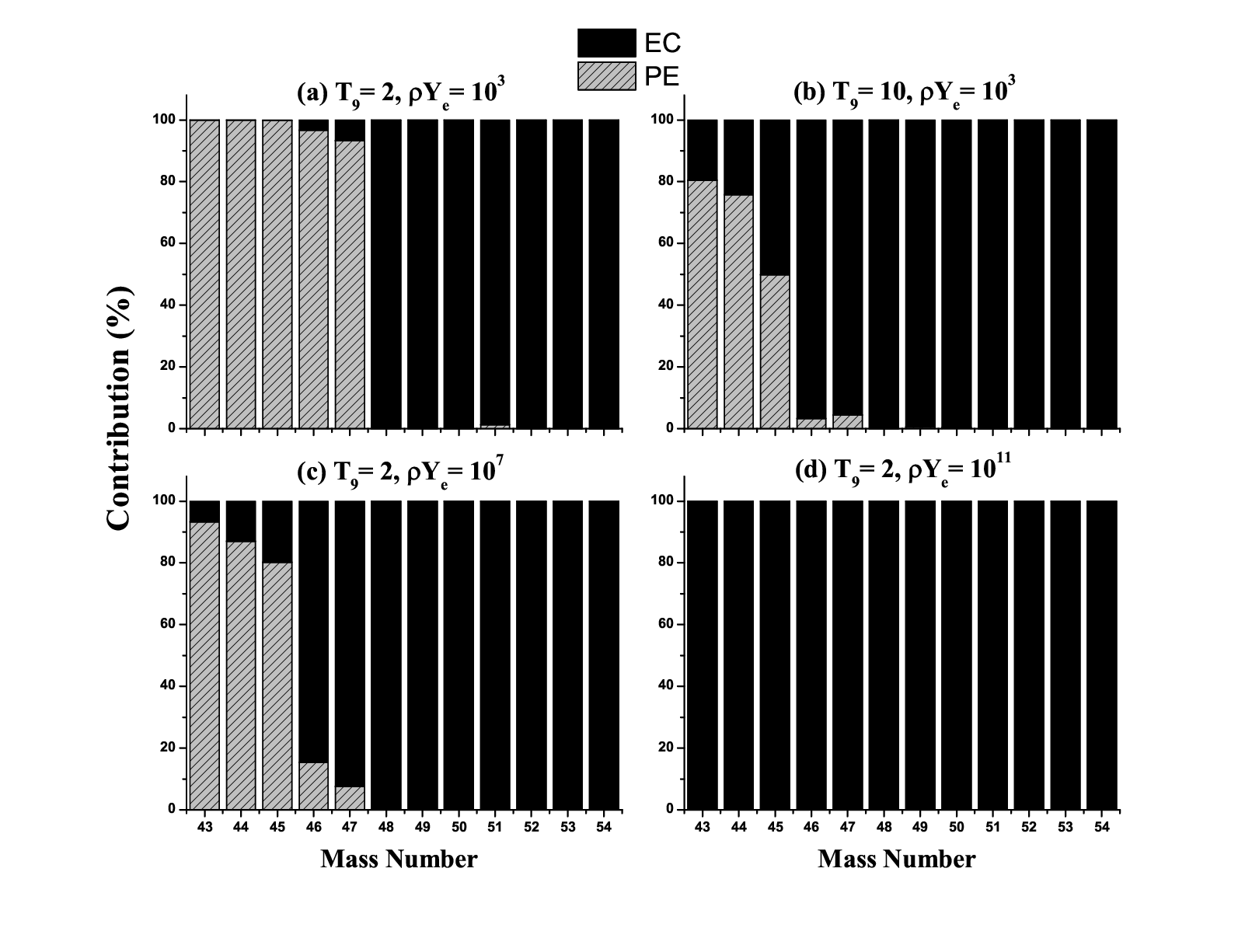}
\vspace{-0.1cm}\caption{Percentage contribution of PE and EC rates to the sum total of weak rates in $\beta^{+}$ direction. Other details are same as in Figure~\ref{EE_PC_Cont}.}
\label{PE_EC_Cont}
\end{center}
\end{figure}

The positron capture (PC) and EC rates act in the same directions as the EE and PE rates, respectively, in changing
the Y$_{e}$ of the stellar core. At times, these LC rates tend to compete
with the corresponding LE rates. Figures~\ref{EE_PC_Cont} (\ref{PE_EC_Cont})
show the percentage contribution of EE (PE) and PC (EC) rates of V-isotopes
to the sum total of weak rates. The percentage contributions are calculated
at different values of stellar temperatures (T$_{9}$=2, 10, 30$\;$K) and
densities ($\rho$Y$_{e}= 10^{3}, 10^{7}, 10^{11}\;$g/cm$^{3}$). From
Figure~\ref{EE_PC_Cont}, it can be observed that for most of the isotopes
under study ($^{49-64}$V), there is negligible contribution of PC rates to
the overall rates in $\beta^{-}$ direction at low temperature and high
density. At high temperature (T$_{9}$ = 30) and low density ($\rho$Y$_{e}=
10^{3}$), the PC rates contribution to the total rates is large, however for
neutron-abundant nuclei, the contribution of EE rates become significant at
high-density ($\rho$Y$_{e}= 10^{11}$). In case of PE and EC rates contribution
to the total weak rates (see Figure~\ref{PE_EC_Cont}), for $^{43-47}$V, PE
rates compete with EC rates in low to medium density regions at low
temperatures. However, at high temperatures and density, EC rates take lead
over PE rates in $\beta^{+}$ direction.

\begin{figure}
\begin{center}
\includegraphics[width=0.8\textwidth]{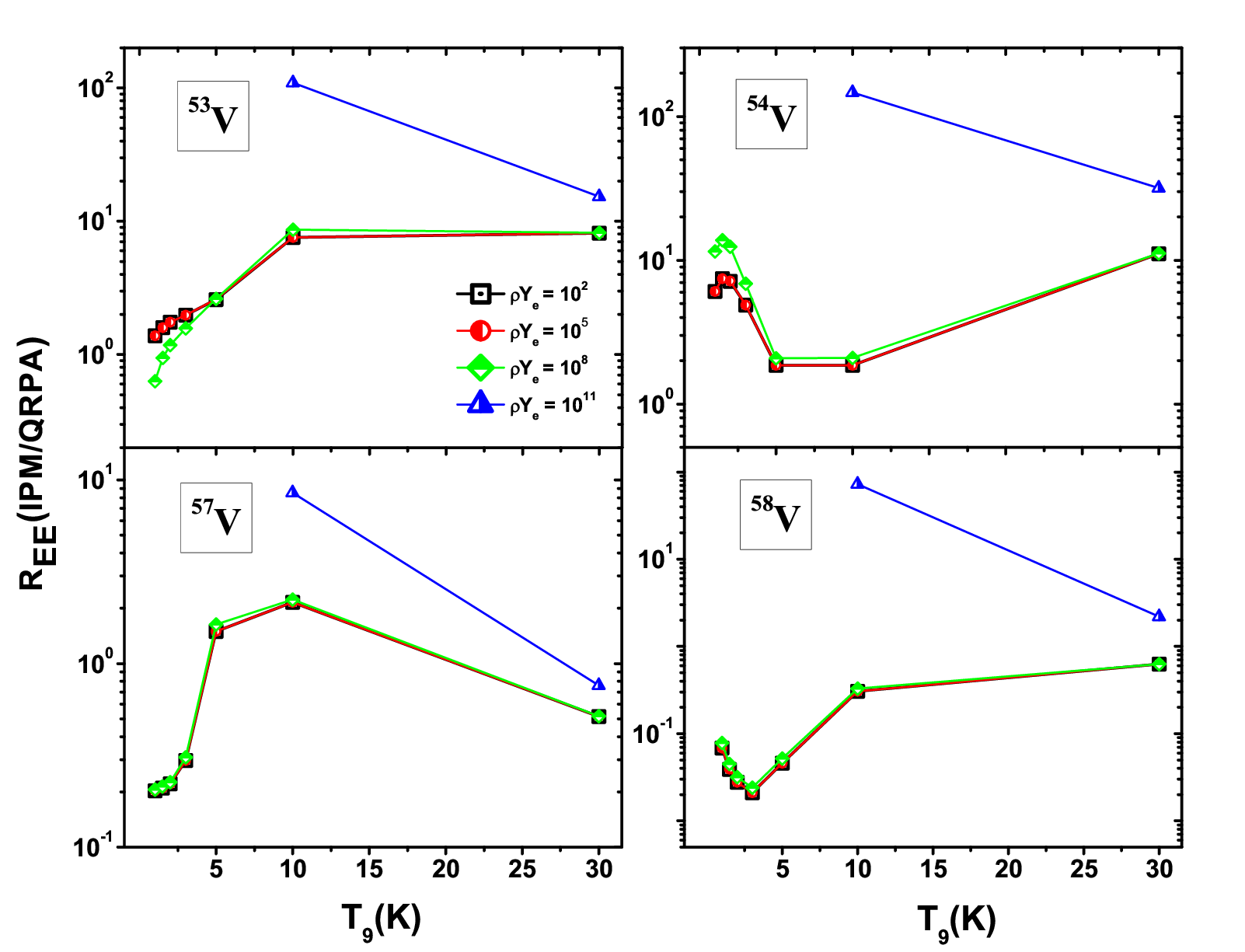}
\includegraphics[width=0.8\textwidth]{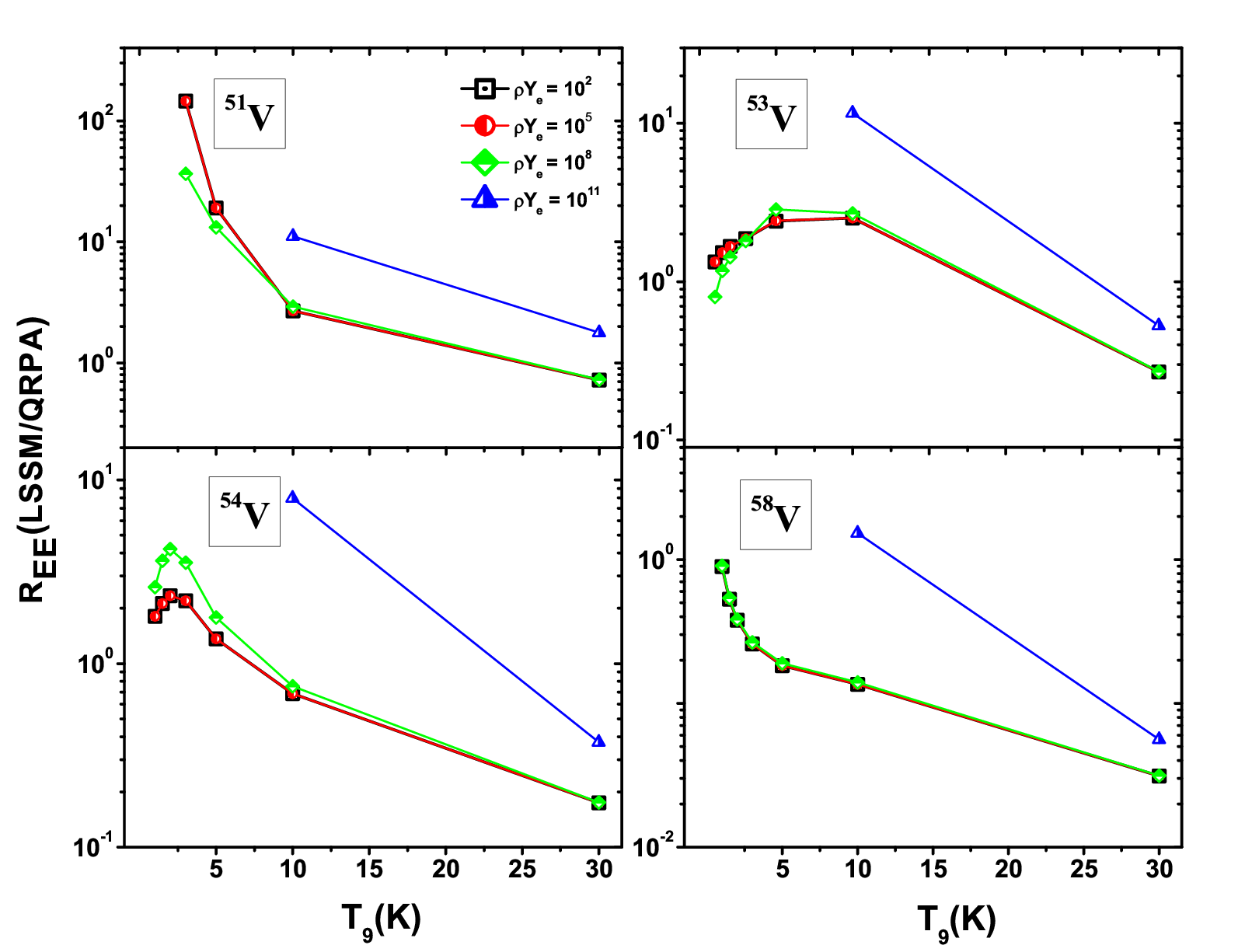}
\vspace{-0.1cm}\caption{The comparison of the EE rates for some of the selected vanadium isotopes calculated using different models. The ratios of previously calculated IPM (LSSM) rates to those of the pn-QRPA rates are shown at top (bottom). For every isotope, ratios are shown as a function of stellar temperatures, T$_{9}$, at four values of density ($\rho$Y$_{e} = 10^{2}, 10^{5}, 10^{8}, 10^{11}\;$g/cm$^{3}$).}
\label{EE_FFN_QRP_Comp}
\end{center}
\end{figure}

\begin{figure}
\begin{center}
\includegraphics[width=0.8\textwidth]{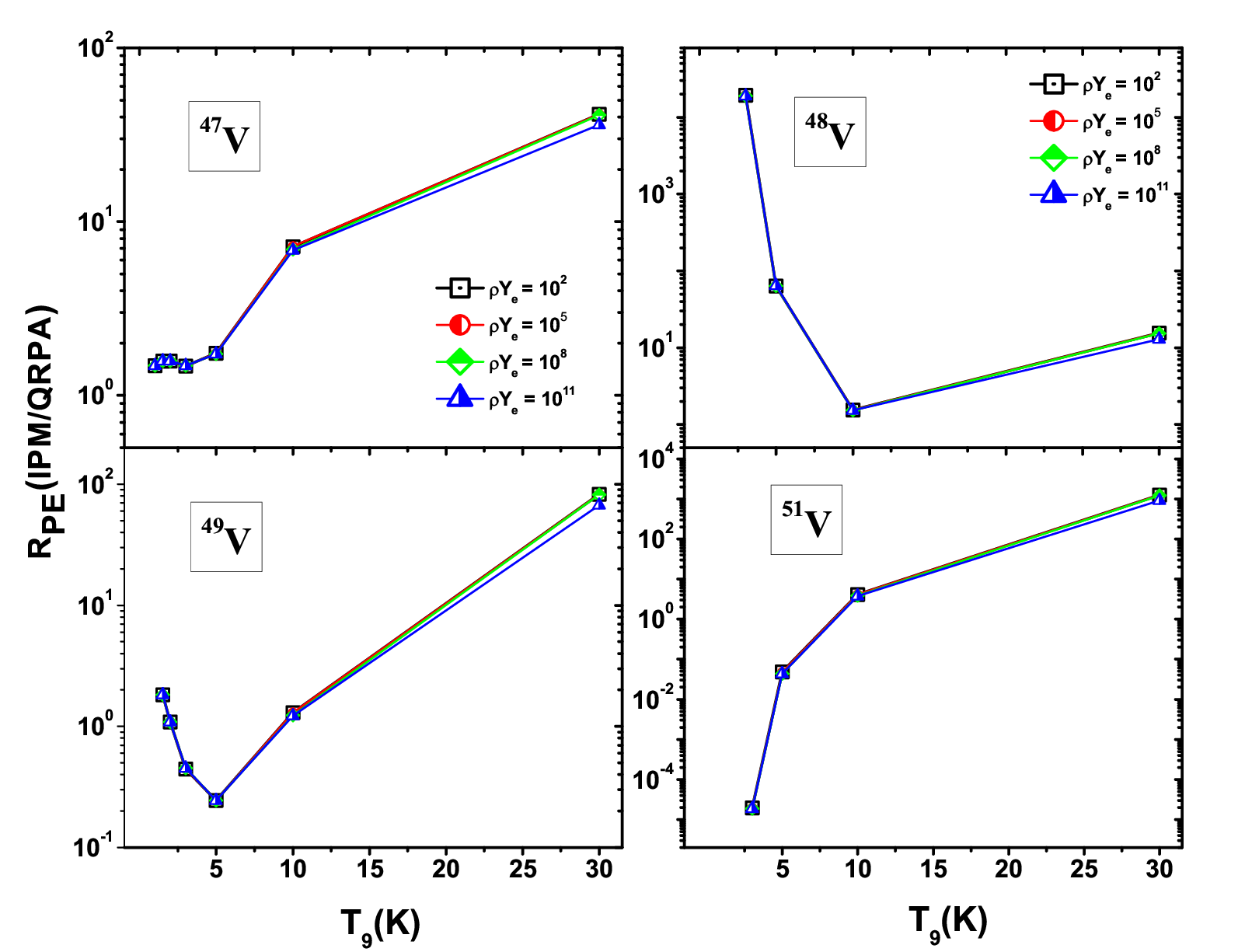}
\includegraphics[width=0.8\textwidth]{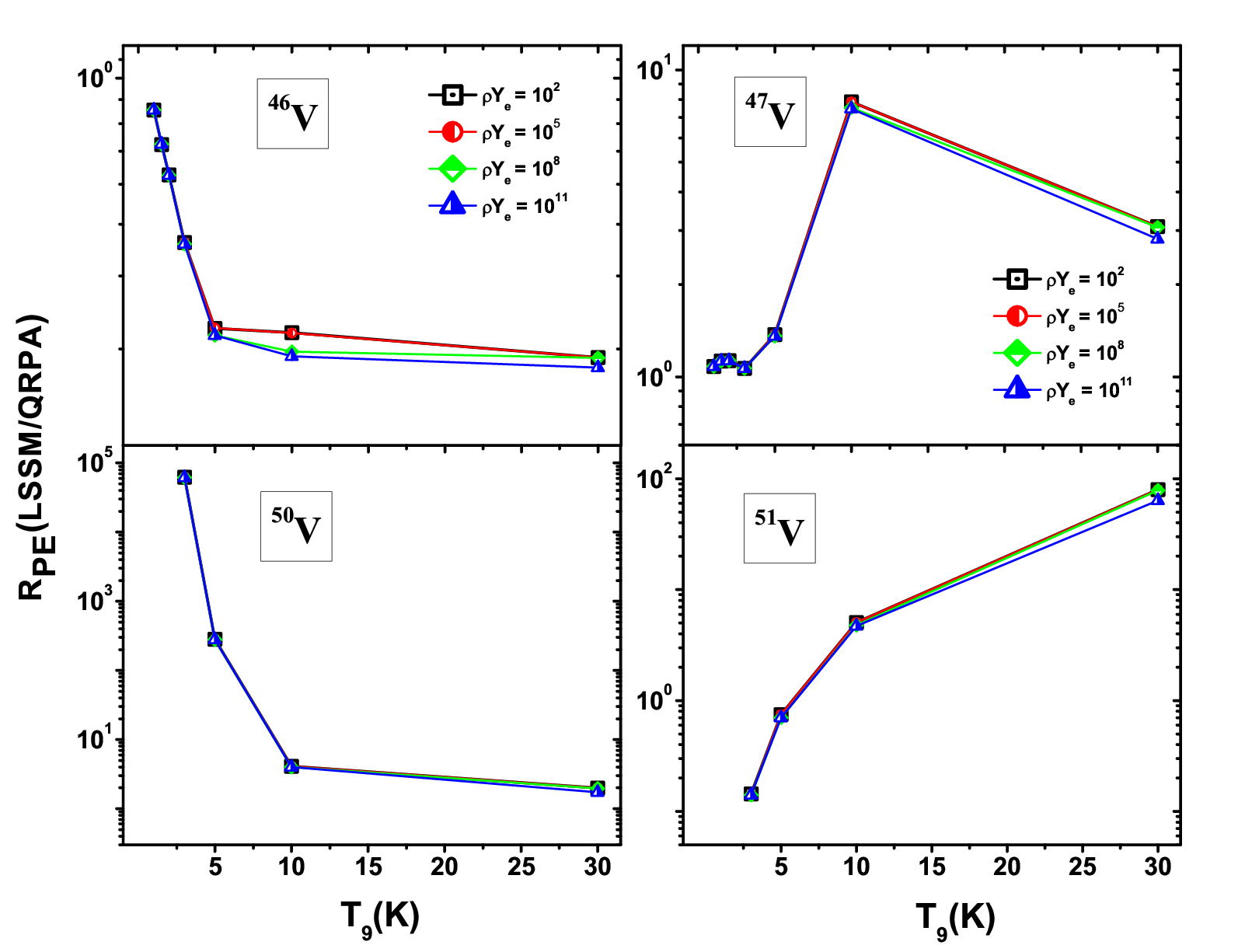}
\vspace{-0.1cm}\caption{The comparison of the PE rates for some of the selected vanadium isotopes calculated using different models. Other details are same as in Figure~\ref{EE_FFN_QRP_Comp}.}
\label{PE_FFN_QRP_Comp}
\end{center}
\end{figure}

Now, we describe the results of comparison of pn-QRPA rates to those of
computed by IPM and LSSM. For the sake of comparison, we have determined the
ratios between our current LE rates and the corresponding IPM and LSSM model
rates, separately. From the calculated ratios, it was observed that various
V-isotopes exhibit a similar trend in their comparison results and hence for
the space consideration, the comparison graphs have been shown only for some
of the selected cases. The graphs of these ratios have been presented in
Figures~\ref{EE_FFN_QRP_Comp} and \ref{PE_FFN_QRP_Comp}. The above set of
four panels in these figures show the ratios of our estimated EE/PE rates to the rates of IPM
and the bottom four panels represent ratios with corresponding LSSM rates. In
each figure, the ratios are plotted against temperature under stellar
conditions at four different values of density ($\rho Y_{e}=$10$^{2}$,
10$^{5}$, 10$^{8}$ and 10$^{11}$ g/cm$^{3}$).

In Figure~\ref{EE_FFN_QRP_Comp}, the top four panels present the comparison
ratios of IPM and our calculated EE rates for two odd-A ($^{53,57}$V) and
two-oven-A ($^{54,58}$V) isotopes. In case of $^{47,48,50,51,54,55}$V, in
general the pn-QRPA rates are lower by some factor to 4 orders of magnitude
by the corresponding IPM rates. Figure~\ref{EE_FFN_QRP_Comp} presents one
such case ($^{54}$V), where it can be observed that in low to medium density
region, at all temperatures our EE rates are smaller in magnitude than IPM
rates by factor $\sim$ 2 to 11. At high density ($\rho Y_{e}=$10$^{11}$
g/cm$^{3}$), for T$_{9}\geq~$10$\;$K, our rates are even smaller by 1-2
orders of magnitude. For $^{53,56}$V, at low temperatures
(T$_{9}\leq$5$\;$K) for densities up to 10$^{8}$ g/cm$^{3}$, the current EE
rates are roughly equal to those of the IPM rates. However at
T$_{9}\geq~$10$\;$K and also at higher density, present rates are reduced by a
factor of $\sim7$ to 1-2 orders of magnitude. The reason for this reduction
in our rates could be attributed to the fact that in the calculations of IPM
the quenching of GT strength was not considered. In addition, the IPM
calculations did not consider the emission of particle from the excited
levels and the excitation energies of parent states are larger than the
particle decay energies. The cumulative effect of these higher excited states
starts to increase at higher values of temperature and density. Therefore under these physical conditions,
the IPM rates are enlarged by 1-2 orders of magnitude.
In case of $^{57}$V,  EE rates from both (IPM and QRPA) models are comparable
within a factor of 5. In lower and medium density regions and for the whole
temperature domain, our estimated EE rates on $^{58}$V are enhanced by that
of IPM model by factor of about two to an order of magnitude. At high
temperatures and density, again the IPM rates are greater than ours for
earlier mentioned reasons. A similar behaviour has been observed for
$^{49,52}$V isotopes. This unusual reduction in IPM rates at $\rho Y_{e} \leq
10^{8}$ may occur because of the assignment of approximated values (which is
too small) to the unmeasured matrix elements in their calculations.

Now we discuss the lower four panels of Figure~\ref{EE_FFN_QRP_Comp}, which
depict the ratio of LSSM model EE rates to that of our model. Overall, in the
set of nuclide $^{47-51}$V, the LSSM rates are larger than our calculated EE
rates by some factors to 3-4 orders of magnitude. One such case ($^{51}$V) is
shown Figure~\ref{EE_FFN_QRP_Comp}. The top left graph in bottom set of
panels shows that, for $^{51}$V in almost whole density and temperature
domain LSSM calculated rates are bigger than ours by factor $\sim$2 to
two orders of magnitude (except at T$_{9}=$30, where the rates from
two models are comparable). In case of $^{53,55}$V our EE rates have good
agreement with that of LSSM within a factor 4-6. Also in case of
$^{52,54,56,57,58}$V, at low temperatures both models rates are nearly equal to each other within
2-4 factors. At higher temperatures (T$_{9}\geq$10$\;$K), however,
our EE rates are increased by some factors to an order
of magnitude than LSSM rates (e.g., see cases of $^{54}$V and $^{58}$V in the
lower graphs of Figure~\ref{EE_FFN_QRP_Comp}). The overall formalism of both
theories; LSSM and pn-QRPA, for the computations of phase space and Q-values
appears the same. However, the employment of Brink's hypothesis and
back-resonances in the theory of LSSM may lead to the above stated differences in
the EE rates estimations in contrast to the pn-QRPA model which
microscopically deals with GT strength calculation of the excited states.

Lastly, we discuss the comparison of PE rates between our pn-QRPA model and
other IPM and LSSM models, separately. This comparison is presented in
Figure~\ref{PE_FFN_QRP_Comp} in the form of ratios in top (bottom) four
panels for IPM (LSSM) models. On average, in the set of nuclei
$^{45,46,47,49}$V, for the whole density domain at low temperatures (T$_{9}
\lesssim 5$), both models rates are in reasonable agreement within factor
$\sim$2-4. However, at higher temperatures, IPM model rates are larger in
magnitude than ours by several factors to roughly an order of magnitude (see e.g., results for $^{47}$V and $^{49}$V in Figure~\ref{PE_FFN_QRP_Comp}). This
IPM rates enhancement in high temperature and density regions can be
attributed to the same reasons as mentioned in case of EE rates. In case of
$^{48,50,52,53}$V, overall the rates of IPM model are bigger than ours by
factor $\sim$2 to 4 orders of magnitude. Figure~\ref{PE_FFN_QRP_Comp}
shows one such example ($^{48}$V), where it can be observed that in all
temperature and density domains our rates are smaller than IPM rates.
Larger differences in the rates of the two models appear at low temperatures
where the PE rates are in itself small. Also in case of $^{51}$V our PE rates
at high temperatures (T$_{9}=$10, 30$\;$K) are reduced by up to $\sim$3 orders
of magnitude then those of IMP rates, however at low temperatures where the
rates are small, enhancement in our rates is observed.

The LSSM and pn-QRPA PE rates comparison shows that, for $^{45,46}$V, overall
our PE rates are bigger in contrast to the LSSM rates by factor $\sim$2-5 (see e.g the
case of $^{46}$V in Figure~\ref{PE_FFN_QRP_Comp}). For $^{47,49,51}$V, at
higher temperatures (T$_{9}=$10, 30$\;$K) LSSM rates get bigger by some factor
to an order of magnitude. At lower temperature, either the rates from both
models show reasonable agreement (see e.g., $^{47}$V) or pn-QRPA rates are
increased by factor $\sim$2-7 (see e.g., $^{51}$V in
Figure~\ref{PE_FFN_QRP_Comp}). In case of three isotopes $^{50,52,53}$V, in
general LSSM PE rates are larger by factor $\sim$2 to 2-4 orders of
magnitude. From this set, $^{50}$V case is shown
Figure~\ref{PE_FFN_QRP_Comp}. The larger difference is observed at lower
temperatures where the rates are small. The differences in the
PE rates from pn-QRPA and LSSM models can again be assigned to the reasons
already discussed.

\section{Conclusions} \label{sec:conclusions}
Presently, we have studied the impact of weak lepton emission processes in
astrophysical environment. For this purpose, a series of vanadium isotopes
with mass numbers in the range A=43 to A=64 was considered. The lepton
emission rates were computed for this set of nuclei considering wide domains
of temperature ($10^{7}- 3\times10^{11}$) K and density ($10^{1}-10^{11}$)
g/cm$^{3}$, under astrophysical conditions. For the calculations of these
rates, the values of GT strength distributions were taken from our previously
published work~\citep{shehzadi20}. Our calculated PE rates for
neutron-deficit V-isotopes ($^{47-49}$V) are bigger than corresponding EE
rates. For $^{50}$V and $^{51}$V, from low to medium density domain, the EE
rates become comparable to the corresponding PE rates. For neutron-rich
nuclei ($^{52-54}$V), EE rates are bigger in magnitude than PE rates on these
nuclei for densities, $\rho Y_{e}<10^{11}$. Although, at high density $\rho
Y_{e}=10^{11}$ for $^{50-54}$V, EE rates are still insignificant as
compared to PE rates.

Present study also incorporates the comparison of pn-QRPA calculated lepton emission rates with the earlier results of corresponding
emission rates of IPM and LSSM. In the comparison of our EE/PE rates with those of IPM, it can be observed that in majority of cases, in the high-temperature and density regime,
IPM rates are larger than our rates. Likewise, PE rates of LSSM are greater than ours in many cases, especially at high temperature.
In contrast, in case of EE, majority cases are those in which our model rates are nearly equal to corresponding rates of LSSM
having good agreement with them. This scenario can be observed at low and high temperature. A basic probable reason of
variations between our and their rates is the use of Brink hypothesis in their models. While, our model gets contributions from excited states
microscopically. A bigger model space approaching to 7$\hbar\omega$ was considered in our pn-QRPA computations.
Other causes which may contribute to these difference are not applying the GT strength quenching, 0$\hbar \omega$ shell-model
estimations of GT centroids, and approximated nuclear matrix-elements in IPM and back-resonances employed in LSSM.

\section*{Acknowledgements}
J.-U. Nabi would like to
acknowledge the support of the Higher Education Commission Pakistan
through project numbers 5557/KPK\\/NRPU/R$\&$D/HEC/2016.

\end{document}